\documentclass{iopart}
\usepackage{graphicx}
\usepackage{cite}
\usepackage{iopams}
\newcommand\ii{\mathrm{i}}

\begin{document}

\title[Chirality for three wave guides]
{Chiral behaviour for three wave guides at an exceptional point of third order}

\author{  W D Heiss$^{1,2}$, G Wunner$^{3}$,}

\address{$^1$Department of Physics, University of Stellenbosch,
  7602 Matieland, South Africa}
\address{$^2$National Institute for Theoretical Physics (NITheP), Western Cape,\
  South Africa}
\address{$^3$Institut f\"ur Theoretische Physik, Universit\"at Stuttgart,
  Pfaffenwaldring 57, 70\,569 Stuttgart, Germany}
\ead{dieter@physics.sun.za}

\ead{wunner@itp1.uni-stuttgart.de}
\begin{abstract}
A simple matrix model that has been used to describe essential features of a 
$\mathcal {PT}$ symmetric set-up
of three coupled wave guides is investigated. The emphasis of the study lies on the occurrence of an exceptional
point of third order. It is demonstrated that the eigenfunctions in close vicinity of the exceptional point have a
distinctive chiral behaviour. Using data describing realistic situations it is argued that such chiral behaviour
can be tested experimentally.
\end{abstract}
 
\submitto{\JPA}
\maketitle

\section{\label{sec:intro} Introduction}
Exceptional points are positions in the parameter space of a physical or
mathematical problem where two or more eigenvalues become degenerate {\it and} the
corresponding eigenvectors coalesce. They can occur only for non-Hermitian operators. 
Exceptional points are of specific importance in
parity-time ($\mathcal {PT}$) symmetric systems, i.\,e. systems described
by non-Hermitian Hamiltonians which are invariant under the combined
parity and time-reversal operation, and which for some parameter ranges can possess
real eigenvalues \cite{Bender98}. It is in this context that exceptional points of
second order (EP2) have been investigated in recent years in a variety of physical
systems, both theoretically and experimentally (for references 
see, e.\,g., \cite{Moiseyev2011a, Heiss12, Bender13,schnabel2017}).

Exceptional points of higher order have also been studied in the literature, even
though mostly theoretically \cite{Graefe08a, Heiss08, 
Graefe12a, Ryu12, Gutoehrlein13, Teimourpour2014, Heiss2015,Am-Shallem15,
Am-Shallem15b,Ding15,Jing2016}. In \cite{Graefe12a} the suggestion was made
that exceptional points of third order (EP3) could be realized experimentally in a
$\mathcal{PT}$-symmetric arrangement of three coupled optical wave guides with
balanced gain and loss in the outer wells. The analysis, however, was restricted
to an {\it ad hoc} mathematical three-dimensional matrix model.

More recently it has been shown \cite{schnabel2017}, by solving 
the full Helmholtz  equations governing three coupled wave guides including gain and loss
under realistic experimental conditions,
that indeed an EP3 can be realized in practice by appropriately
tuning the complex index of refraction profiles of the wave guides. Moreover, it was
also shown \cite{schnabel2017a} that for these realistic parameters a matrix of the type discussed in
\cite{Graefe12a} is recovered by projecting the full equations on the subspace 
spanned by the three ground state modes of the wave guides.

EP3s are of special interest for a variety of reasons. In particular, they exhibit, in contrast to EP2s,  chiral
behaviour in {\it the eigenstates} \cite{Heiss08}. When one moves away from the EP3 in parameter space 
three eigenvalues emerge, forming an equilateral triangle to lowest order in the perturbation. The
phase relations of the associated eigenstates depend on the orientation
of the triangle. With the above mentioned results \cite{schnabel2017} the experimental
test of this chiral signature seems to become possible.

In the present paper we study the chiral behaviour near the EP3 in the 
$\mathcal{PT}$-symmetric three wave guide system. In view of the fact that for 
realistic experimental parameters a matrix of the type discussed in \cite{Graefe12a}
was established, we confine ourselves
to study the chiral signature using the simpler matrix model.

\section{The Matrix Model for three Wave Guides}
\label{model}
In a recent paper \cite{schnabel2017} a realistic model of three wave guides has been investigated with
particular focus upon the occurrence of an exceptional point of third order (EP3).
For the numerical work the quoted paper is based on an analytic solution of a Helmholtz equation.
In a preceding MS-thesis \cite{schnabel2017a} the same problem has been reduced to a three-dimensional matrix problem
where the matrix coincides with the matrix introduced by \cite{Graefe12a} for the same problem. It reads
\begin{equation}
mat=\begin{pmatrix} {a+2 \ii g & \sqrt{2} v & 0 \cr  \sqrt{2} v & 0 &  \sqrt{2} v  \cr
        0 &  \sqrt{2} v & a-2 \ii g }
\end{pmatrix}
\label{mtrx}
\end{equation}
where the real parameter $g$ denotes the gain and loss, the real $v$ the coupling and $a$ is a perturbation
of the complex refractive index in the outer wells that can be used to encircle the EP3. Actually, the gain/loss term
is introduced in \cite{schnabel2017} by $\pm \ii g\lambda /(2\pi )$ with $\lambda=1.55~{\mu m}$. Thus, every parameter
in the matrix has the dimension inverse length and is given in units of $(\mu m)^{-1}$.

The (complex) eigenvalues of $mat$ are denoted by $k$.
The matrix has an EP3 at $k=0$ for $g=v$ and $a=0$. The eigenfunction at the EP3 -- a coalescence of three 
eigenfunctions -- is up to a factor,
\begin{equation}
|\psi_{\rm EP3}\rangle =\begin{pmatrix} {\ii \cr \sqrt{2} \cr -\ii}.
\end{pmatrix}
\end{equation}
Note that the $c$-norm of $\psi_{\rm EP3}$ vanishes \cite{Heiss01,Moiseyev2011a}.
For the convenience of the reader we here give a short rehash of the properties of the eigenfunctions
in the vicinity of an EP3 \cite{Heiss08}. 

Due to the cubic root character of the EP3 a generic perturbation of one particular parameter causes three
eigenvalues (here of $mat$) to sprout off the EP3. To lowest order in the perturbing parameter they form
an equilateral triangle. The corresponding eigenvectors form a complete eigensystem
in the three-dimensional space. The three coefficients of an expansion of $\psi_{\rm EP3}$ in terms of this
orthonormal system have ratios $(1,\exp (2\ii \pi/3),\exp(- 2\ii \pi/3))$ or $(1,\exp(- 2\ii \pi/3),\exp (2\ii \pi/3))$
depending on the specific orientation of the associated eigenvalues. Each of the two cases specifies a
particular orientation and therefore chiral behaviour. Depending on the particular problem, there may be
more than one or specific combinations of parameters to invoke the three different eigenvalues under perturbation.
In the simple matrix example above a (complex) perturbation of $a=0$ produces three eigenvalues.

To obtain such result in general we stress that a judicious choice of the perturbing parameter or combinations must be made.
As discussed in \cite{schnabel2017} there can be perturbations with different parameters which invoke
two EP2s to pop out from the EP3. In fact, an EP3 can be seen as the confluence
of two EP2s that share a Riemann sheet. To invoke this type of behaviour usually a specific combination of
parameters must be used.

This discussion clearly manifests the rich structure associated in general with an EP3. The advantage of the simple
matrix model introduced here lies in the considerable reduction of the complicated structures associated with
an EP3 in more general cases. It suffices to demonstrate the chiral behaviour of eigenfunctions near to the EP3.

\begin{figure}
  \centering
  \includegraphics[width=0.4\columnwidth]{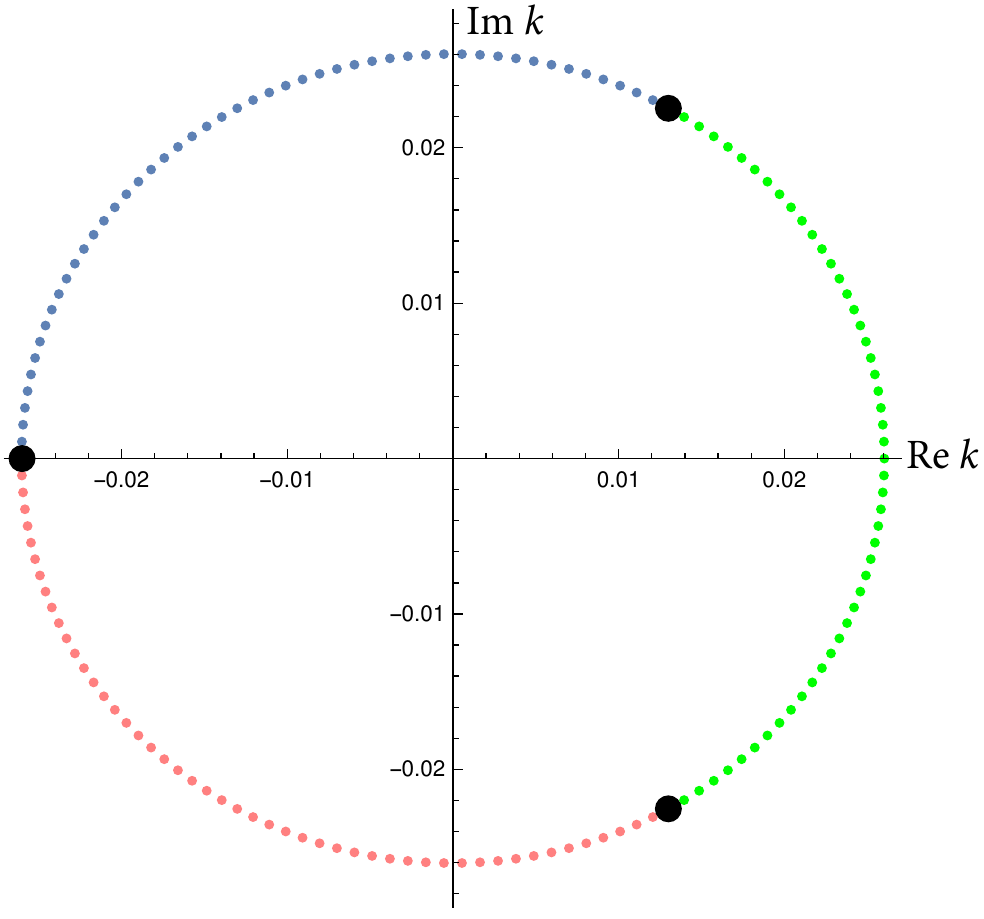}
  \includegraphics[width=0.4\columnwidth]{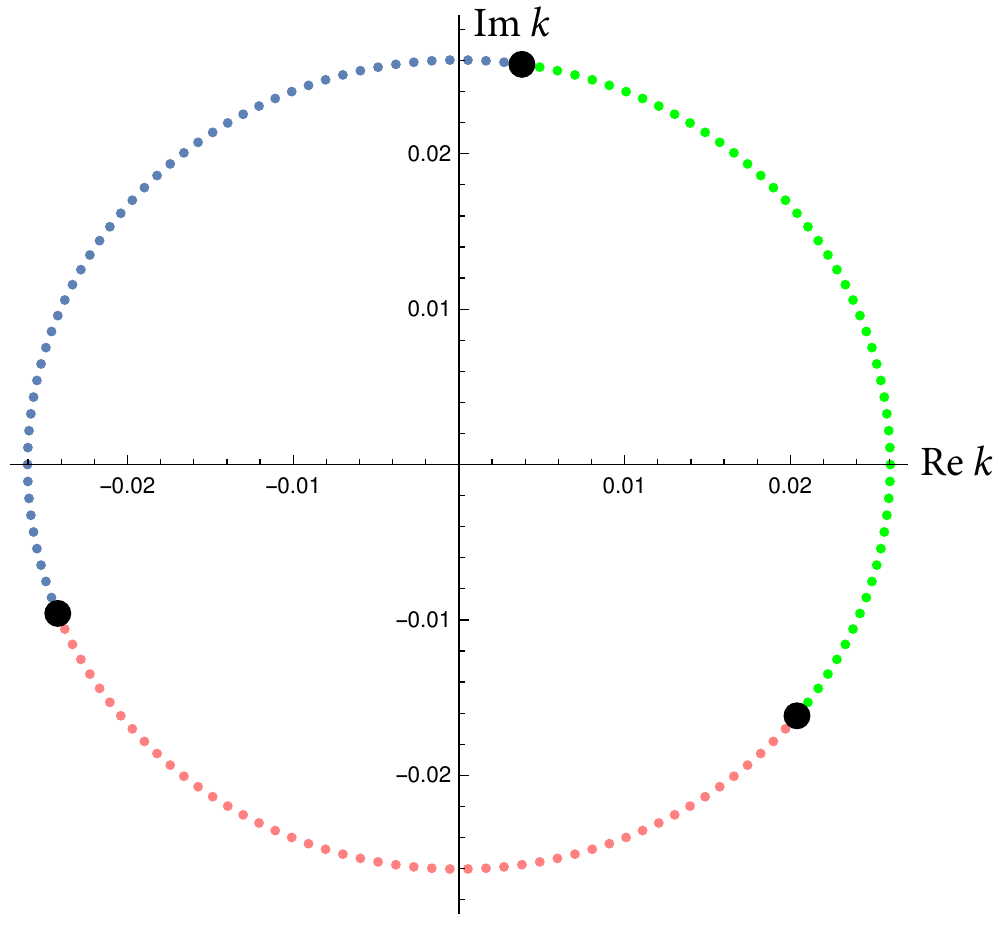}
  \caption{Left: Values in the complex $k-$plane when the parameter $a$ describes three times a circle around the EP3 (at $a=0$)
  in the complex $a-$plane. The black dots indicate the respective starting point at $a=10^{-6}$. The illustration
  is for $g=v=2.1$. For smaller values of $g=v$ the figure remains qualitatively the same but the values are smaller.\\
  Right: Same as the left figure but with the starting point at $a=10^{-6} \exp(2\ii \pi 9/50)$.}
  \label{fig:three}
\end{figure}

\section{Numerical Results and Discussion}
\subsection{Eigenvalues}
The EP3 at $k=0$ for $a=0$ is obtained for a wide range of values for $g=v$. In Fig.1 we illustrate the complex $k-$plane
for three full rounds in dependence of exact circles in the complex $a-$plane. 
The left figure indicates the respective three eigenvalues
at the starting point $a=10^{-6}$ and displays the respective images in the $k-$plane for the values
$a=10^{-6}\exp(2\ii \pi m/50)$ for $m=0,\ldots,50$. Qualitatively similar figures are obtained for values of $g=v$ in the range
$2\ge g \ge 10^{-4}$ when the circle in the $a-$plane is kept small. An increase of the radius of the circle in the
$a-$plane displays, however, structures from further singularities as mentioned qualitatively in the preceding section.
We illustrate examples in Fig.2. Realistic values for $g$ are even smaller 
\cite{schnabel2017} such as about $10^{-6}$.
In the following we focus upon the chirality of the eigenfunctions. 

Before turning to the eigenfunctions we have to specify the ranges of the two basic positions of the three
eigenvalues that characterize the two possible chiralities. If we consider the eigenvalues of Fig.1 but with starting values
at $a=10^{-6}\exp(2\ii \pi m/50)$ and $m$ running from 1 to 24 for each full circle 
(or we may think of the whole continuum in between), then the starting points form triangles
that are turned by the angle $0^0<\alpha <60^0$ with the effect that two eigenvalues lie below the real axis. 
An example is given on the right-hand side of Fig.1 using the arbitrary value $m=9$. In turn,
considering the eigenvalues for the starting points $a=10^{-6}\exp(-2\ii \pi m/50)$ with the same $m-$values the corresponding
triangles are turned into the other
direction in such a way that two eigenvalues lie above the real axis. It is for these two ranges where the associated
eigenvectors have the one or the other chirality as shown in the following subsection. When an eigenvalue lies exactly on
the real axis as is the case when $m$ is 0 or 25, a definite chirality cannot be attributed to this state. Recall
that for this to hold we always consider a deviation to first order, i.e.~sufficiently small values of $a$.

\begin{figure}
  \centering
  \includegraphics[width=0.4\columnwidth]{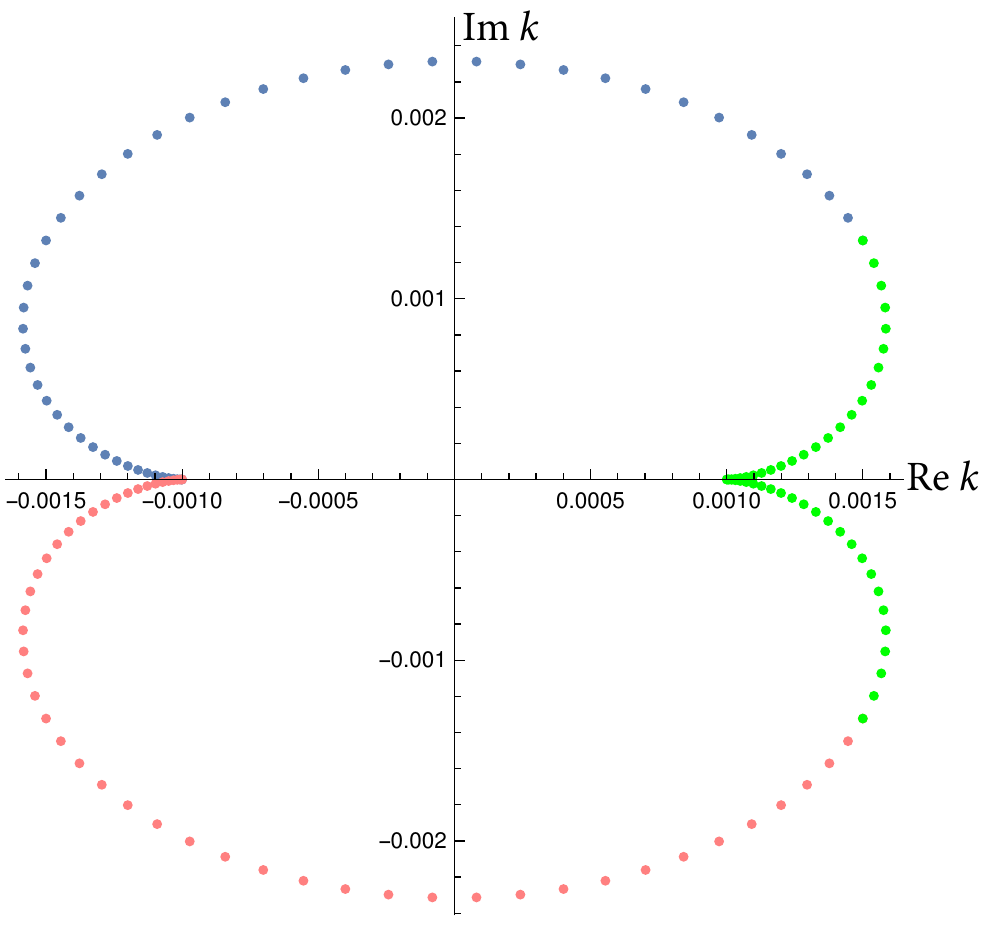}
   \includegraphics[width=0.4\columnwidth]{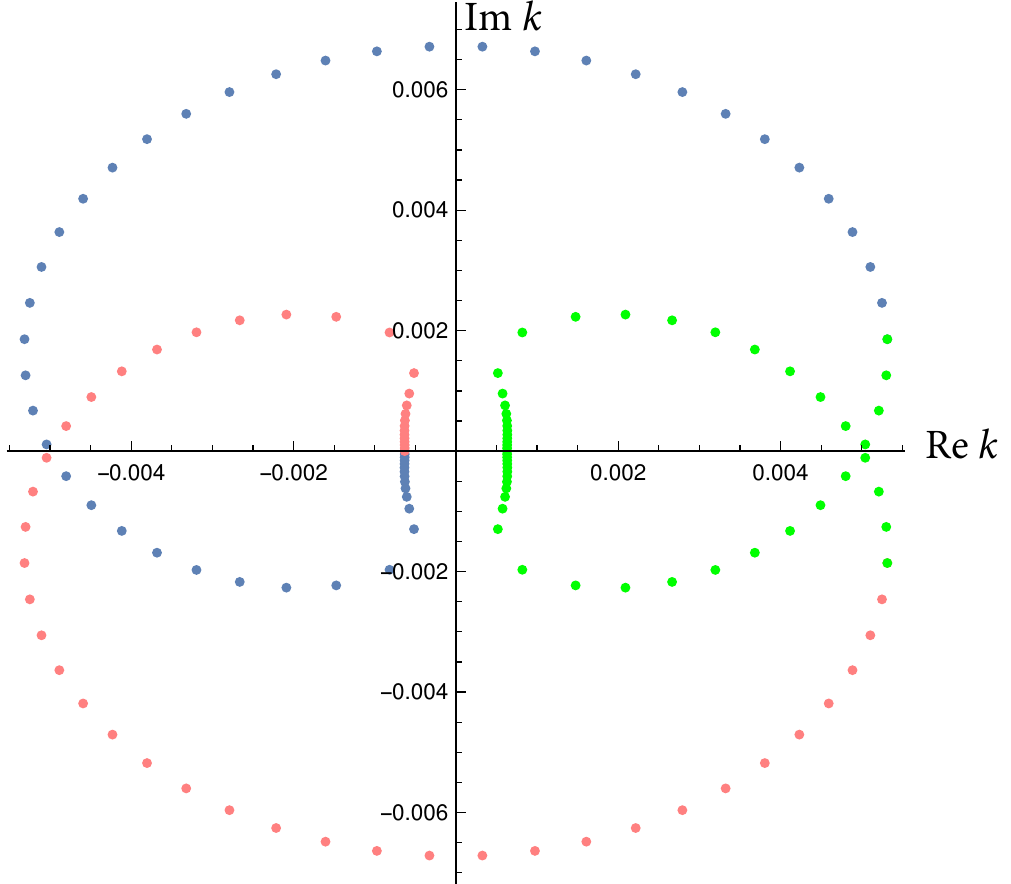}
  \caption{Left: Similar to Fig.1 but with the larger radius $a=0.001$ and smaller values $g=v=0.001$.\\
          Right: Same  but with the larger radius $a=0.005$. The effects of further singularities nearby (EP2)
         are clearly discernible }
  \label{fig:smaller_a}
\end{figure}

\subsection{Eigenvectors}
Except at the EP3 the matrix $mat$ of Eq.(\ref{mtrx}) has eigenvectors that span the three-dimensional space. We denote them by
$|\psi_i(a)\rangle $, where we indicate the dependence on $a$, and we assume them to be normalized by the
$c-$norm $\langle \tilde \psi_i(a)|\psi_j(a)\rangle =\delta_{ij}$ with $\langle \tilde \psi_i(a)|$ being a left-hand
eigenvector of $mat$. Note that the $c-$norm vanishes at the EP3, i.e.~for $a\to 0$. In other words, at the EP3
there is no (obvious)\footnote {We don't consider associate vectors in this context.}  
complete system, the three eigenfunctions coalesce. For $a\ne 0$ we may expand $\psi_{\rm EP3}$
\begin{equation}
|\psi_{\rm EP3}\rangle =\sum_i C_i(a)|\psi _i(a)\rangle .
\label{exp}
\end{equation}

Note that this expansion holds identically in $a\ne 0$. Note further that, while the normalized functions 
$|\psi _i(a)\rangle $ blow up for $a\to 0$ the coefficients  $C_i(a)$ vanish accordingly to yield the finite
left-hand side of Eq.(\ref {exp}). As a consequence of the vanishing $c-$norm of $|\psi_{\rm EP3}\rangle $ we also have
$$ \sum_i C_i^2(a)=0.  $$
The essential point \cite{Heiss08} are the ratios of the coefficients in close vicinity of the EP3, {\it viz.}
\begin{equation}
 \pmatrix{C_1(a)\cr C_2(a)\cr  C_3(a)}=\xi(a) \pmatrix{1\cr \exp (2\ii \pi/3)\cr  \exp (-2\ii \pi/3)}
\label{c1}
\end{equation}
or 
\begin{equation}
 \pmatrix{C_1(a)\cr C_2(a)\cr  C_3(a)} =\eta(a) \pmatrix{1\cr \exp (-2\ii \pi/3)\cr  \exp (2\ii \pi/3)}
\label{c2}
\end{equation}
when the corresponding eigenvectors $|\psi_i(a)\rangle$ are numbered according to the ascending order of their associated
real parts of their eigenvalue. Here
$\xi(a)$ and $\eta(a)$ are some complex numbers that vanish to lowest order as $\root 3 \of {|a|}$
for $a\to 0$.
Whether (\ref{c1}) or  (\ref{c2}) applies
depends on the ordered orientation of the triangle formed by the eigenvalues. 

As discussed in the previous subsection we may turn the triangle formed by the black dots in Fig.1
by an angle $\alpha $ in the range $0^0<\alpha <60^0$
counter clockwise or clockwise. In the former case the relative phases of $ C_i(a)$ follow relation  (\ref{c2})
and in the latter case relation (\ref{c1}). We recall that this result is valid only to lowest order in $a$, if $a$
is chosen too large the relation is no longer valid (as would be the case for respective eigenvalues in Fig.2).

We believe that this distinctive chiral behaviour is amenable to measurement. In fact, similar phase differences
have been observed for the eigenfunctions at an EP2 \cite{Stehmann04, Dietz2011}, but there is no chiral behaviour. Here,
the eigenfunctions at the EP3 have no specific phase behaviour while the first order perturbation yields our result.
What is encouraging for an experimental verification is our finding that even for very small values of $g=v$, -- as 
they appear to be in a realistic model \cite{schnabel2017} --, an accordingly small value for $a$ still yields the pattern as
in Fig.1; in other words, our result appears to be valid for realistic input.

\subsection{Time evolution}
To discern the different phases in the three eigenstates the stationary states have to be compared.
When time goes by, it will of course always be the state with the smallest width that survives.
However, for intermediate times a specific beat from the interference of the respective phase differences
should be discernible as long as the eigenvalues are sufficiently close to each other, i.e.~to the EP3. 
We recall that at the singularity no signal is expected \cite{HeissWunner16,Klaiman08a}.

\section{\label{concl}Summary and outlook}
For a simple matrix model a specific chiral behaviour of the eigenstates in the vicinity of an EP3
has been demonstrated depending on the geometric orientation of the three eigenvalues. 
Since the model bears the essential structure of an EP3 the result should be valid
for any situation giving rise to an EP3 if a perturbation by a suitable parameter gives rise to the sprouting
out of three distinct eigenvalues from the EP3. For convenience we have chosen a $\cal{ PT-}$symmetric problem
that allows the EP3 to occur at a real value. Nevertheless, the results are also valid for three resonances
when they form an EP3 as the matrix used in (\ref{mtrx}) can be modified by adding a negative imaginary
part in the diagonal elements that shifts all eigenvalues into the lower $k-$plane.

We emphasise that, in contrast to the EP2, the chiral behaviour is present just in the vicinity of the EP3. Basically,
it is a result of the analytic behaviour of a third root. We are aware of recent results for EP2 where
chiral behaviour is established by dynamically circumscribing an EP2 \cite {Doppler2016a,Xu2016a}. This is a different phenomenon
that is expected to have an analogy for EP3; in view of the richer structure of an EP3 this aspect requires
a more involved investigation that goes beyond the scope of this paper. As we have implemented our results numerically
with realistic values for three wave guides in the microwave region we are confident that an experimental confirmation
should be feasible.

\section*{Acknowledgements}
WDH and GW gratefully acknowledge the support from the National Institute for Theoretical Physics (NITheP), 
Western Cape, South Africa. GW expresses his gratitude to the Department of Physics of the University of
Stellenbosch where this work was carried out. The authors thank Mr.~J.~Schnabel for making available his numerical data of
the matrix model.

\section*{References}

\end{document}